\documentclass{PoS}
\usepackage{amsmath}
\renewcommand{\Re}{{\rm Re\thinspace}}

\newcommand{\Lumint}{{\cal L}_{\rm int}}

\title{Charged Higgs boson benchmarks \\
  in the CP-violating type-II 2HDM}

\ShortTitle{Charged Higgs boson benchmarks in the CP-violating type-II 2HDM}

\author{Lorenzo BASSO$^a$, Anna LIPNIACKA$^b$, Farvah MAHMOUDI$^{cd}$, Stefano~MORETTI$^{ef}$, Per OSLAND$^b$, \speaker{Giovanni Marco PRUNA}\thanks{This work has been supported by the German Research Foundation DFG through Grant No.\ STO876/2-1 and by BMBF Grant No.\ 05H09ODE.}$^{\ g}$ and Mahdi~PURMOHAMMADI$^b$ \\ \ \\
\llap{$^a$} Albert-Ludwigs-Universit\"at - Fakult\"at f\"ur Mathematik und Physik, \\ 
        D-79104 Freiburg i.\ Br., Germany \\ \ \\
\llap{$^b$} Department of Physics and Technology, University of Bergen, \\
        Postboks 7803, N-5020 Bergen, Norway\\ \ \\
\llap{$^c$} CERN Theory Division, Physics Department, \\
        CH-1211 Geneva 23, Switzerland \\ \ \\
\llap{$^d$} Clermont Universit{\'e}, Universit\'e Blaise Pascal, CNRS/IN2P3, LPC, \\ 
        BP 10448, 63000 Clermont-Ferrand, France\\ \ \\
\llap{$^e$} School of Physics \& Astronomy, University of Southampton, Highfield, \\
        Southampton SO17 1BJ, UK \\ \ \\
\llap{$^f$} Particle Physics Department, Rutherford Appleton Laboratory, \\
        Chilton, Didcot, Oxon OX11 0QX, UK \\ \ \\
\llap{$^g$}
        TU Dresden, Institut f\"ur Kern- und Teilchenphysik, \\
Zellescher Weg 19, D-01069 Dresden, Germany\\ \ \\
        E-mails: \email{Lorenzo.Basso@physik.uni-freiburg.de}, \email{Anna.Lipniacka@ift.uib.no}, \email{Mahmoudi@in2p3.fr}, \email{S.Moretti@soton.ac.uk}, \email{Per.Osland@ift.uib.no}, \email{Giovanni\_Marco.Pruna@tu-dresden.de}, \email{Mahdi.PurMohammadi@ift.uib.no}}

\abstract{
We discuss prospects for discovering a charged Higgs boson in the CP-violating type-II two-Higgs-doublet Model.
In the light of recent LHC data, we focus on the parameter space that survives experimental as well as the theoretical constraints on the model.
Once the phenomenological scenario is set, we analyse the scope for the LHC to discover a charged Higgs in association with the lightest neutral Higgs and a charged vector boson.
}

\FullConference{Prospects for Charged Higgs Discovery at Colliders\\
		 October 8-11, 2012\\
		 Uppsala University Sweden}

\begin{document}

\section{Introduction}

The recent discovery of the $\sim 125$ GeV resonance at the LHC \cite{:2012gk,:2012gu} leads the searches for physics Beyond the Standard Model (BSM) to a crucial point: despite its high compatibility with the SM prediction reinforces the robustness of the latter, still we know that the SM is an effective theory which is missing particular ingredients such as neutrino masses, gravity, etc. (e.g., see \cite{Djouadi:2005gi} for details). In this regard, the LHC is the only chance to explore concrete BSM terascale extensions in the near future. Among the options, one popular set is represented by the supersymmetric (SUSY) scenarios, and much effort has been devoted in this direction by the physics community. However, in the less favourable situation, the scale of energy at which this kind of theories can be fully probed could be well beyond the LHC capability. If this occurs then the only portal for ``indirect'' SUSY detection is through the profiling of the Higgs sector. 

Moreover, the main feature of any SUSY potential is to have at least two Higgs doublets set in the so-called type-II configuration (e.g., see \cite{Djouadi:2005gj}), and the peculiarity of a two-Higgs-doublet model (2HDM) is that a charged scalar is contained in the particle spectrum \cite{Gunion:1989we}. Therefore, the discovery of a charged Higgs boson would demonstrate the existence of a non-minimal Higgs potential. In a minimal approach, this would be interpreted as the presence of more than one Higgs doublet and possibly (but not necessarily) an underlying SUSY nature of the Universe. Furthermore, it is known \cite{Gorbahn:2009pp} that quantum corrections will lead to the mixing between scalars and a pseudoscalar in a SUSY tree-level potential, and this mechanism will produce an effective CP-violating potential. Following these motivations, we consider the scope for the LHC to discover a charged Higgs which stems from an explicitly CP-violating 2HDM with type-II Yukawa couplings, in association with the lightest neutral Higgs and a charged vector boson, with the former decaying into the lightest neutral Higgs and a second $W$ boson. The final state altogether yields a $b\overline{b} W^+W^-$ signature, of which we exploit the $W^+W^-$ semileptonic decays~\cite{Basso:2012st}.

\section{The Model}

In this Section, we describe the parametrisation of the CP-violating 2HDM with Type-II Yukawa couplings. 
The Higgs sector is defined by the presence of two Higgs doublets, with one ($\Phi_2$) coupled to the $u$-type quarks, and the other ($\Phi_1$) to the $d$-type quarks and charged leptons, in analogy with the Minimal Supersymmetric Standard Model (MSSM).

Following \cite{Ginzburg:2002wt,Khater:2003wq}, we take the scalar potential to be
\begin{eqnarray}
\label{Eq:pot_7}
V&=&\frac{\lambda_1}{2}(\Phi_1^\dagger\Phi_1)^2
+\frac{\lambda_2}{2}(\Phi_2^\dagger\Phi_2)^2
+\lambda_3(\Phi_1^\dagger\Phi_1) (\Phi_2^\dagger\Phi_2) \nonumber \\
&+&\lambda_4(\Phi_1^\dagger\Phi_2) (\Phi_2^\dagger\Phi_1)
+\frac{1}{2}\left[\lambda_5(\Phi_1^\dagger\Phi_2)^2+{\rm h.c.}\right] \\
&-&\frac{1}{2}\left\{m_{11}^2(\Phi_1^\dagger\Phi_1)
\!+\!\left[m_{12}^2 (\Phi_1^\dagger\Phi_2)\!+\!{\rm h.c.}\right]
\!+\!m_{22}^2(\Phi_2^\dagger\Phi_2)\right\}, \nonumber
\end{eqnarray}
where $\lambda_5$ and $m_{12}^2$ are complex parameters which trigger the mixing between CP-violating and CP-conserving components. Following the parametrisation of \cite{ElKaffas:2006nt,ElKaffas:2007rq}, we start from $8$ degrees of freedom: $3$ of them are the would-be Goldstone components of the massive gauge bosons, then we are left with $5$ Higgs bosons:  $H_1$, $H_2$, $H_3$ and $H^\pm$. Then, the whole parameter space is described by $8$ parameters: $\tan\beta$ (ratio of VEVs), $\sin{\alpha_i}$ (mixing parameters, with $i=1,3$), $M_1$ (mass of $H_1$), $M_2$ (mass of $H_2$), $M_{H^\pm}$ (mass of $H^\pm$), $\mu$ (combination of Lagrangian mass parameters and VEVs: $\mu^2=\Re m_{12}^2/(2\cos\beta\sin\beta)$).

We remark that CP violation potentially modifies the structure of any Higgs interaction, hence also the Yukawa couplings are affected. For a complete treatment we invite the reader to consult reference~\cite{Basso:2012st} (and references therein).

\section{The Charged Higgs boson profile}

The multi-dimensional type-II 2HDM parameter space is severely restricted
by a variety of theoretical (\textbf{T}) and experimental (\textbf{E}) constraints, which are
listed in the following:
\begin{itemize}
\item[\textbf{T}:] positivity, tree-level perturbative unitarity, perturbativity, global potential minimum.
\item[\textbf{E}:] $B\to X_s \gamma$, $B_u\to \tau \nu_\tau$, $B\to D\tau \nu_\tau$, $D_s\to \tau \nu_\tau$, $B_{d,s}\to \mu^+\mu^-$, $B^0-\overline{B}^0$ (the SM predictions for the flavour observables are obtained using SuperIso v3.2 \cite{Mahmoudi:2007vz,Mahmoudi:2008tp}), $R_b$, $pp \to H_jX$ (LHC constraints), $T$ and $S$ (EW precision tests), Electron Electric Dipole Moment (EDM).
\end{itemize}

After a detailed analysis of the surviving parameter space, we have defined a set of benchmark points which is collected in Table~\ref{points} \cite{Basso:2012st}, with a preference for points with $\tan\beta\sim \mathcal{O}(1)$.

\begin{table}[!ht]
\begin{center}
\begin{tabular}{|c|c|c|c|c|c|c|}
\hline
\ & $\alpha_1/\pi$ & $\alpha_2/\pi$ & $\alpha_3/\pi$ & $\tan\beta$ & $M_2$ & $M_{H^\pm}^{\text{min}} ,M_{H^\pm}^{\text{max}}$\\
\hline
 $P_1$ & $0.23$ & $0.06$ & $0.005$ & $1$ & $300$ & 300,325\\
 $P_2$ & $0.35$ & $-0.014$ & $0.48$  & $1$ & $300$ & 300,415\\
 $P_3$ & $0.35$ & $-0.015$ & $0.496$ & $1$ & $350$ & 300,450\\
 $P_4$ & $0.35$ & $-0.056$ & $0.43$ & $1$ & $400$ & 300,455\\
 $P_5$ & $0.33$ & $-0.21$ & $0.23$ & $1$ & $450$ & 300,470\\
 $P_6$ & $0.27$ & $-0.26$ & $0.25$ & $1$ & $500$ & 300,340\\
\hline
 $P_7$ & $0.39$ & $-0.07$ & $0.33$ & $2$ & $300$ & 300,405\\
 $P_8$ & $0.34$ & $-0.03$ & $0.11$ & $2$ & $400$ & 300,315\\
\hline
\end{tabular}
\end{center}
\caption{Benchmark points selected from the allowed parameter space when $M_1=125$~GeV. Masses $M_2$ and allowed range of $M_{H^\pm}$ are in GeV, and $\mu$ is set to $200$~GeV.  \label{points}}
\end{table}

\begin{figure}[!t]
\begin{center}
\includegraphics[width=.45\textwidth]{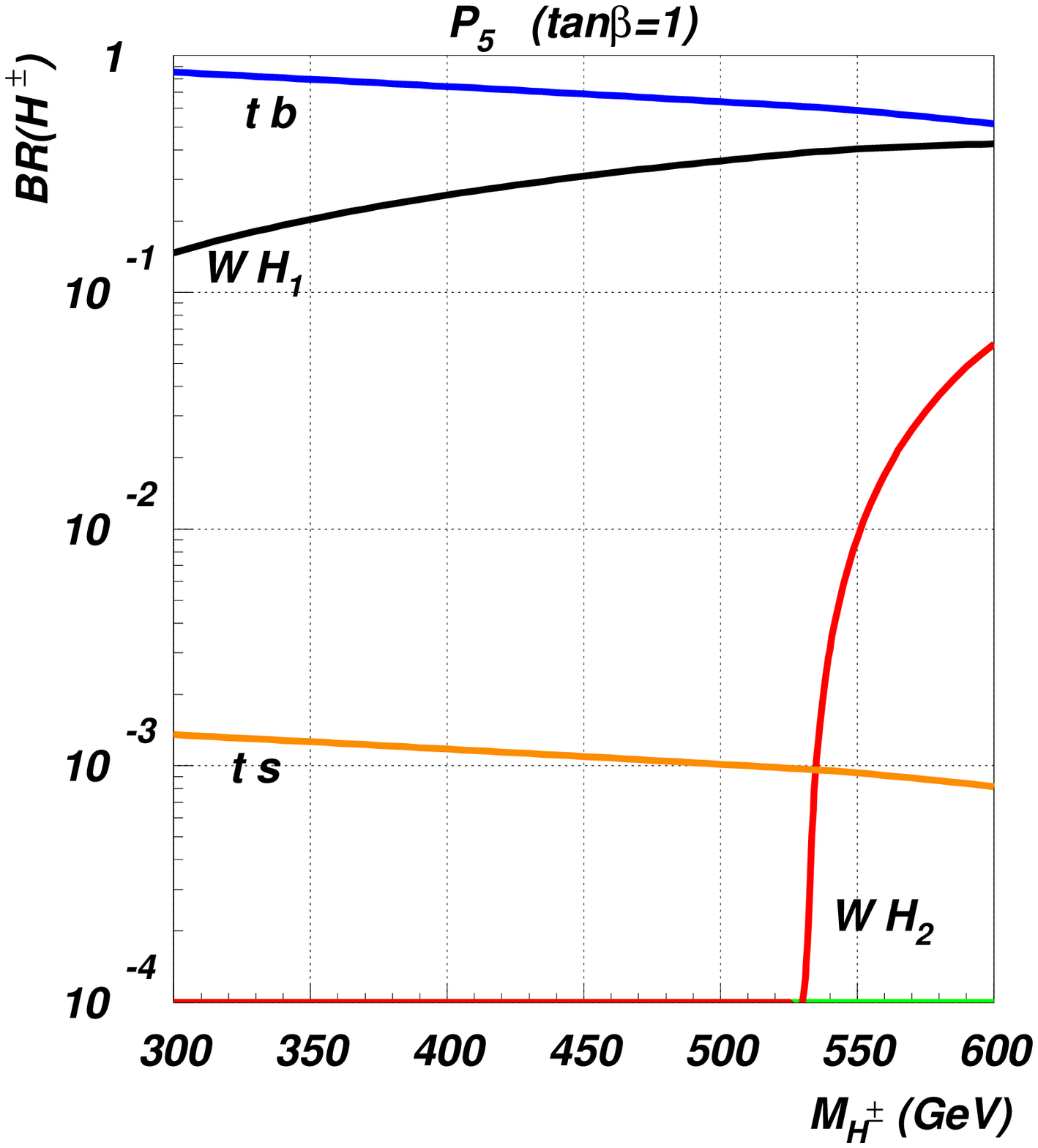} \qquad
\includegraphics[width=.45\textwidth]{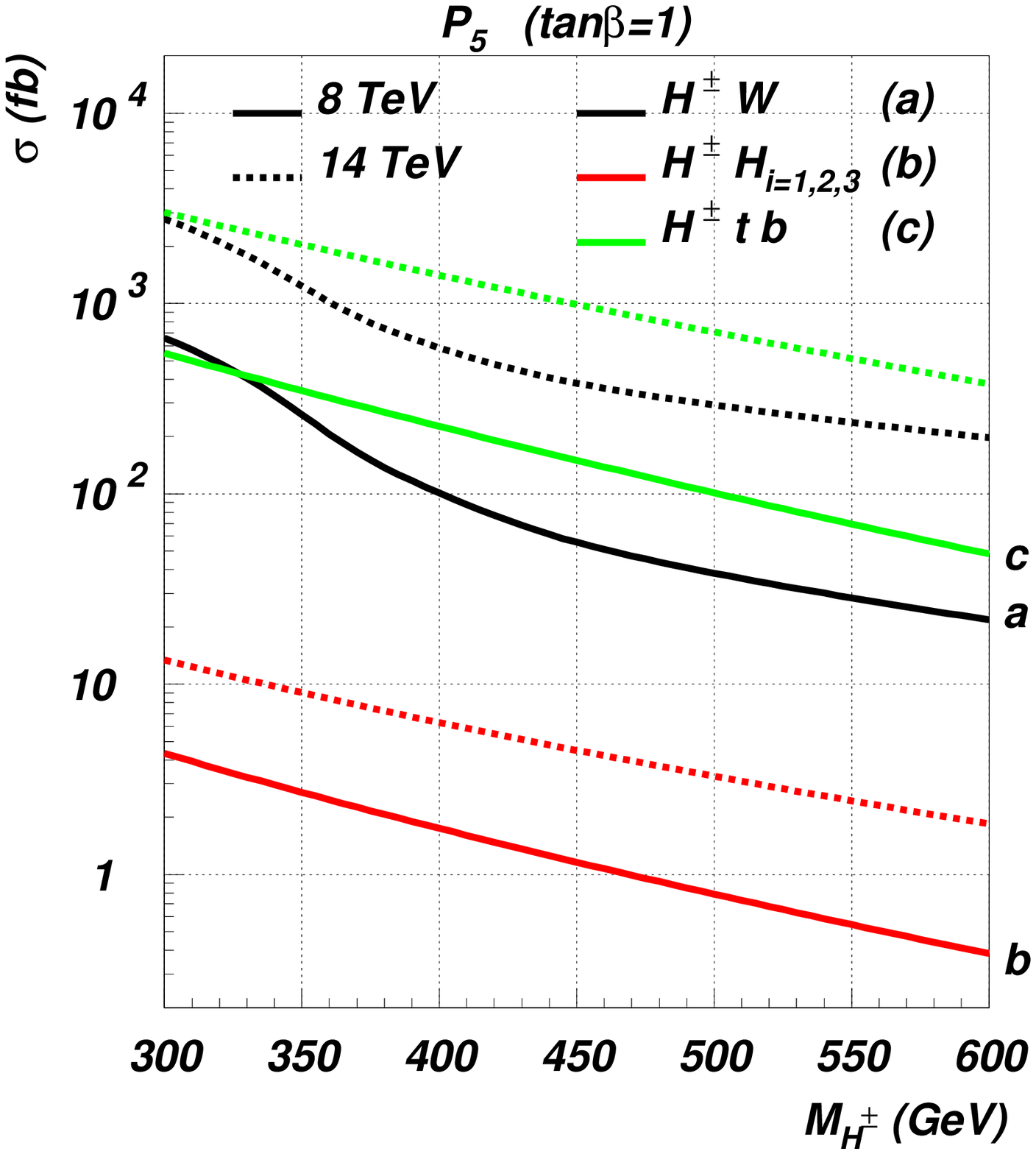}
\end{center}
\caption{BRs (left frame) and production cross section (right frame) of a charged Higgs boson plotted against its mass. The choice of parameters corresponds to benchmark point $P_5$ of Table~1.
\label{fig1}}
\end{figure}

As an example, let us consider $P_5$ and profile the charged Higgs by studying the branching ratios (BR) of the main decay channels and the production cross sections. In order to perform the computation, the model has been implemented via LanHEP \cite{Semenov:2010qt} into CalcHEP \cite{Belyaev:2012qa}\footnote{The implementation of the loop-induced Higgs couplings was made via LoopTools \cite{Hahn:1998yk}. As for the implementation of the Goldstone and ghost sectors of the model see \cite{Mader:2012pm}.}. In the left frame of Figure~\ref{fig1}, we show the BR of the most important charged Higgs decay channels ($tb$, $WH_1$, $ts$ and $WH_2$), from which it is possible to conclude that the decay modes $tb$ and $WH_1$ lie in the range $BR\sim 0.1-1$. From the right frame of Figure~\ref{fig1}, instead, we see that the dominant production mechanisms are via top association ($gg\to t\overline{t}\to H^+b\overline{t}(H^-t\overline{b})$) and gluon fusion ($gg\to H_i\to H^\pm W^\mp$, where $i=1,3$), and for allowed masses of the charged Higgs the cross section is $\sigma\sim \mathcal{O}(10^{2})$ fb. We have verified that this is mostly true for all the benchmarks.

Since we established that the production and decay modes may differ at most by a factor $\sim \mathcal{O}(10)$ and considering the fact that it is very difficult to extract any fermion-associated process from the large $gg\to t\overline{t}$ background at the LHC, we decided to pursue the strategy of analysing the pure boson-associated charged Higgs signal.

\section{Phenomenological strategies}

We studied the process $gg\to H_i\to H^\pm W^\mp$ combined with the decay mode $H^\pm\to W^\pm H_i$, with the neutral Higgs further decaying to $b\overline{b}$-pairs and the $W$ pair taken to decay semi-leptonically (this allows for a full reconstruction of the event and for a final state which is cleaner than the fully hadronic decay mode). We analysed the Gaussian significance ($\Sigma=S/\sqrt{B}$) of such process at the LHC with a hadronic centre-of-mass energy of $\sqrt{s}=14$ TeV and an integrated luminosity of $\Lumint=100$ fb$^{-1}$. For each benchmark point, $2 \cdot 10^4$ unweighted events were produced. Regarding the $t$ background, $4.5 \cdot 10^6$ unweighted events (with generation cuts) have been simulated. For emulating a real LHC-prototype detector, a Gaussian smearing was included to take into account the electromagnetic energy resolution of $0.15/\sqrt{E}$ and the hadronic energy resolution of $0.5/\sqrt{E}$.

We next describe the overall strategy for the background reduction procedure. A first set of  cuts includes typical detector kinematic acceptances and standard intermediate object reconstruction, such as $W\to jj$ and $H_1 \to b\overline{b}$ (cuts 1--3 in \cite{Basso:2012st}). Further, a $t$-(anti)quark reconstruction is used as ``top veto'' (cut~4 in \cite{Basso:2012st}). Guided by the consideration that a $b$ quark pair stemming from the Higgs boson is boosted (unlike the almost back-to-back pair from $t\overline{t}$), we define the last cut of this first set (cut~5 in \cite{Basso:2012st}). After these rather generic cuts are imposed, more signal-based selections can improve the significance.

The main consideration of the following analysis is that the charged Higgs mass can equivalently be reconstructed by either the invariant mass of the four jets ($2b+2j$), $M(b\overline{b}jj)$, or the transverse mass of the $b$ jets, the lepton and the MET, $M_T(b\overline{b}\ell\nu)$. Let's focus on the $M(b\overline{b}jj)$-$M_T(b\overline{b}\ell\nu)$ plane: for the signal, either of the two variables will always reconstruct the correct charged Higgs boson mass, thus producing a cross-like shape in the plane defined by the two masses. In contrast, the background events accumulate at $\sim 2m_t$. The presence of long tails for the signal towards regions where the top background is heavily reduced allows us to introduce a ``single'' cut on the aforementioned plane: 
\begin{eqnarray}\label{cut_sq}
\label{cut_sg}
\mbox{``single cut'': }\qquad \mbox{C}_{\mbox{sng}} &=& M_T(b\overline{b}\ell\nu)>M_{\rm{lim}}\, .
\end{eqnarray}
The single cut of Equation~(\ref{cut_sg}) is applied only on $M_T(b\overline{b}\ell\nu)$ because the reduction of the top background is higher than if compared to a similar cut on the $M(b\overline{b}jj)$ for the same numerical value of $M_{\rm{lim}}$.
If a further selection is imposed, restricting the evaluation of the signal-over-background to the peak-region only, i.e.
\begin{equation}\label{cut_peak}
\mbox{peak cut:}\qquad \left|M-M_{H^\pm}\right|<50\mbox{ GeV}\, ,
\end{equation}
then the significance is optimised by imposing $M_{\rm{lim}}=600$ GeV. As for the specific case of $P_5$, when $M_{H^\pm}=310$ (390) GeV the number of surviving events is $\sim 23.3$ (53.4) and the significance is $\Sigma=5.2$ (11.8).

\begin{figure}[!t]
\begin{center}
  \includegraphics[angle=0,width=0.45\textwidth ]{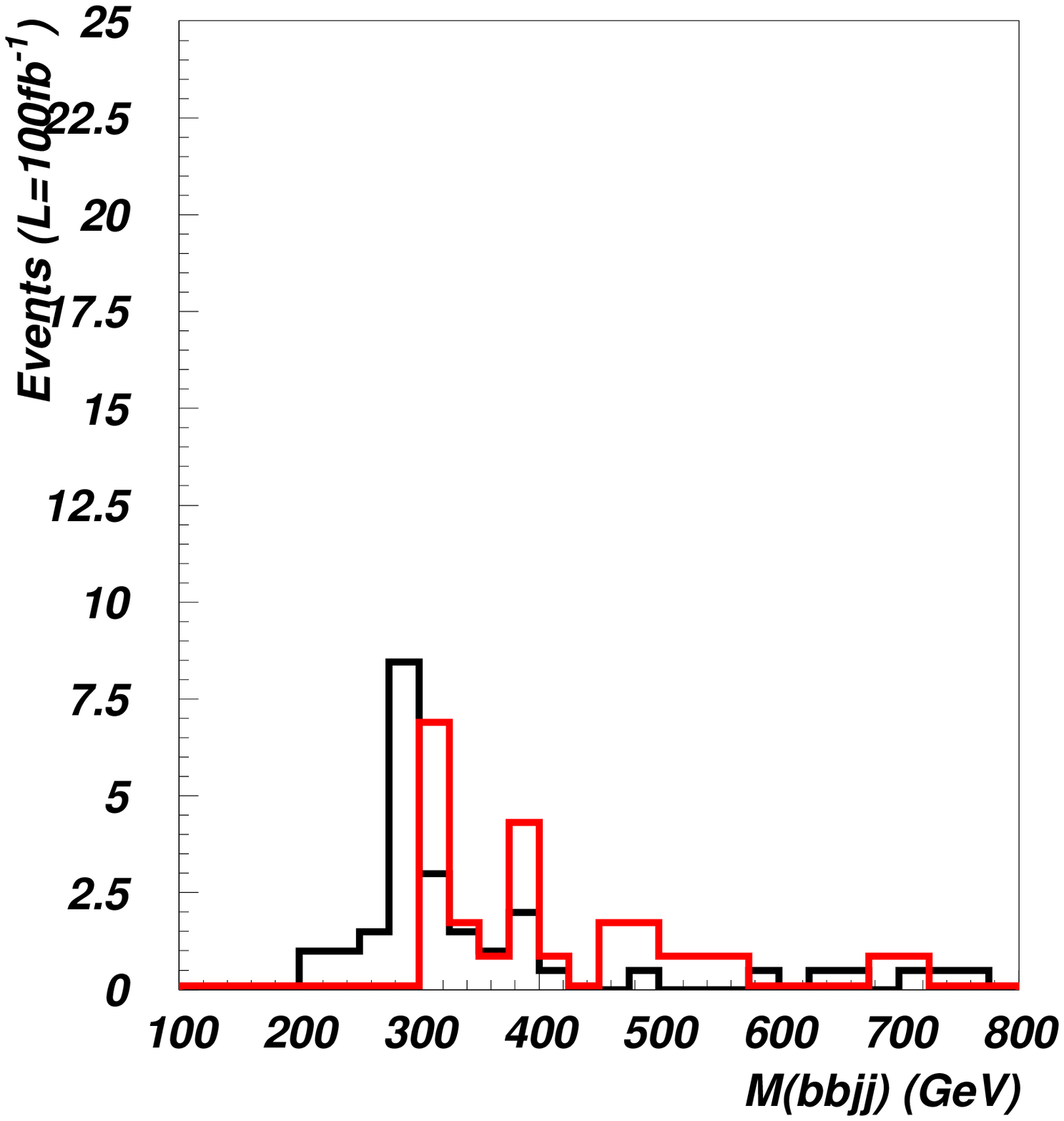} \qquad
  \includegraphics[angle=0,width=0.45\textwidth ]{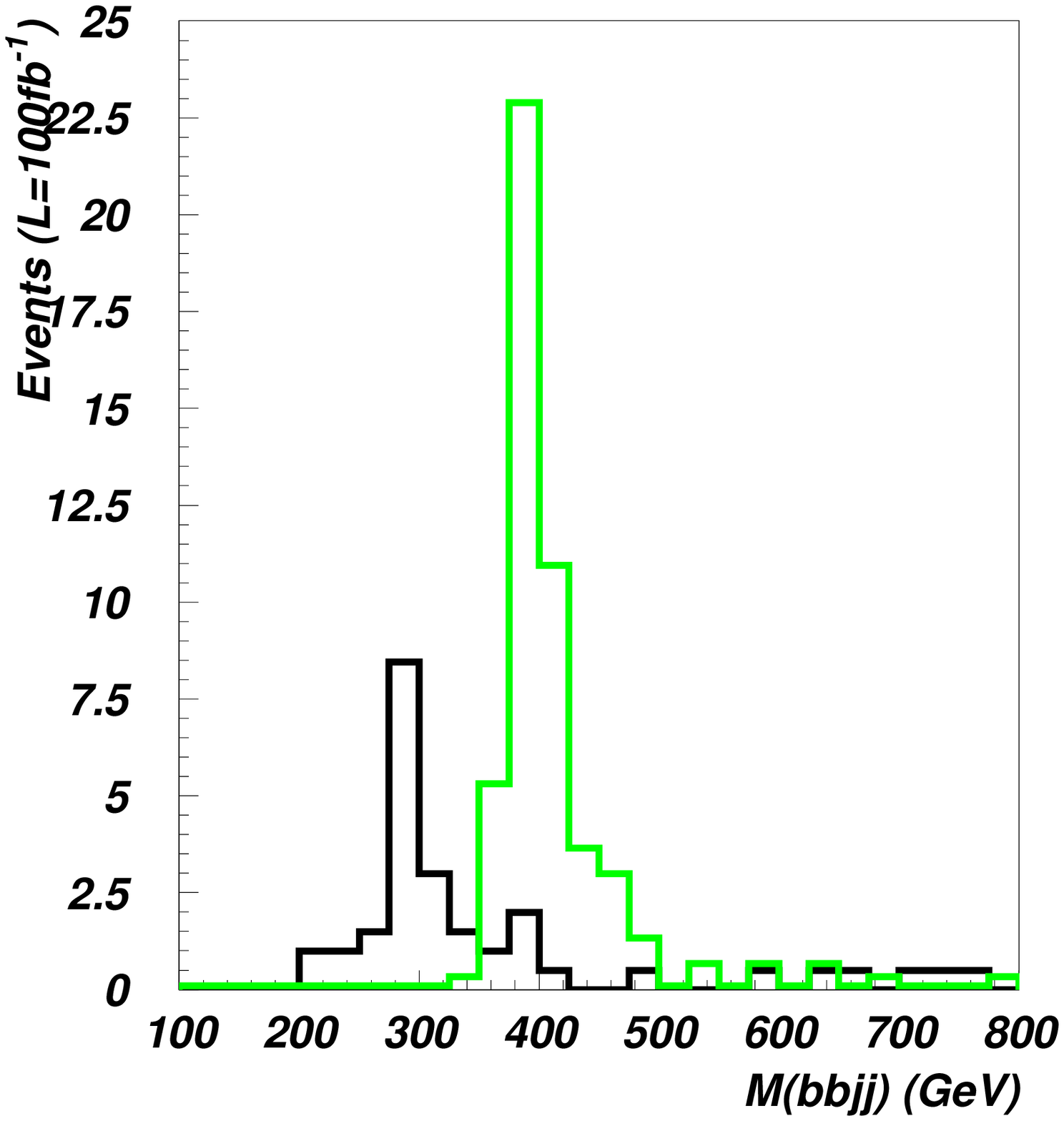} 
\end{center}
\caption{$P_5$. Number of events with $\Lumint=100$ fb$^{-1}$ at $\sqrt{s}=14$ TeV vs $M(b\overline{b}jj)$ for signal (coloured histograms) and $t\overline{t}$-quark background (black). Red (green) corresponds to $M_{H^\pm}=310$ (390) GeV. \label{Pi_5}}
\end{figure}

In Figure~\ref{Pi_5} we plot the number of integrated events as a function of $M(b\overline{b}jj)$ for both the signal and the background with $M_{H^\pm}=310$ GeV (left frame) and $M_{H^\pm}=390$ GeV (right frame), from which we see that the cutting strategy allowed us to extract a signal over the $t\overline{t}$ background.

\section{Conclusion}

We have reviewed a set of benchmarks \cite{Basso:2012st} for the CP-violating 2HDM with Type-II Yukawa interactions to be investigated at the LHC by exploiting the channel $pp\to H^\pm W^\mp\to W^+W^-b\overline{b}$. 
These points all have $M_1=125~\text{GeV}$, 
low $\tan\beta$, they all violate CP, and allow for a range of charged-Higgs masses.

A set of cuts is proposed with the aim of reducing the $t\overline{t}$ background to a tolerable level, and allowing for the detection of a signal in the $WW\to jj\ell\nu$ channel. Many of the proposed benchmark points lead to enhanced $H^\pm W^\mp$ production cross sections due to resonant production via $H_2$ or $H_3$ in the $s$-channel.

Finally, we have shown an explicit example of this successful strategy when applied to the benchmark point $P_5$. In all generality, we conclude that this procedure helps in the isolation of a signal stemming from a charged Higgs boson which is only associated to scalar and vector bosons (both in production and decay).

\end{document}